\documentclass[journal, draftclsnofoot, onecolumn, 12pt]{IEEEtran}
\usepackage{adjustbox}
\usepackage{caption}
\usepackage{fancyhdr}
\usepackage{amsmath, cases, amsfonts, amssymb, txfonts, nccmath, amscd, flushend} 
\usepackage{mathrsfs}
\usepackage{makeidx}
\usepackage{subfigure} 
\usepackage{indentfirst} 
\usepackage{bm} 
\usepackage{multicol}
\usepackage{listings}
\usepackage{color}
\usepackage{booktabs, longtable}
\usepackage{setspace}
\usepackage{hyperref}
\usepackage{booktabs}
\usepackage{stfloats}
\usepackage{setspace, stfloats}
\usepackage{graphicx}
\usepackage{multirow}
\usepackage{array}
\usepackage{tikz}
\usepackage{multicol}
\newenvironment{narrow}[2]{%
\begin{list}{}{%
\setlength{\topsep}{0pt}%
\setlength{\leftmargin}{#1}%
\setlength{\rightmargin}{#2}%
\setlength{\listparindent}{\parindent}%
\setlength{\itemindent}{\parindent}%
\setlength{\parsep}{\parskip}}%
\item[]}{\end{list}}
\newtheorem{rem}{Remark} 

\title{Pattern-Division Random Access (PDRA) for   M2M Communications with Massive MIMO Systems
}
\author{
Xiaoming  Dai,  Tiantian Yan, Qianqian Li, Hua Li  and   Xiyuan Wang \\
\thanks{ Xiaoming  Dai,
Tiantian Yan, Qianqian Li, and Hua Li  are with the School of
Computer and Communication Engineering, University of Science and
Technology Beijing, Beijing 100083, China (e-mail:
 daixiaoming@ustb.edu.cn).}

\thanks{Xiyuan Wang is with the School of Information and Communication
Engineering, Beijing Information Science and Technology
University, Beijing 100101, China (e-mail:
wangxiyuan@bistu.edu.cn).}
}
\begin{document}
\maketitle 
\begin{abstract}
In this work, we introduce the  pattern-domain pilot design paradigm  {based on  a ``\emph{superposition of orthogonal-building-blocks}"}  with significantly larger contention space to enhance the massive machine-type communications (mMTC)  random access (RA)  performance in massive  multiple-input multiple-output (MIMO)  systems.
Specifically,  the pattern-domain pilot is constructed based on  the  superposition of $L$  cyclically-shifted Zadoff-Chu (ZC)  sequences.    The pattern-domain pilots exhibit  zero correlation values between non-colliding patterns from the same root and low  correlation values between patterns from different roots.
  The increased    contention space, i.e., from \emph{N} to $\binom{N}{L}$,  where  $\binom{N}{L}$ denotes the number of all \emph{L}-combinations of a set \emph{N},  and low  correlation values
lead to a significantly lower pilot    collision  probability  without compromising \emph{excessively} on channel estimation  performance for mMTC RA  in   massive MIMO systems.
We present the framework and analysis of the RA success probability of the  pattern-domain  based scheme  with massive MIMO systems.
Numerical results demonstrate  that the proposed pattern division random access (PDRA)  scheme  achieves an appreciable performance gain over the conventional one,
while preserving the existing  physical layer virtually unchanged.   The extension of the ``{superposition of orthogonal-building-blocks}" scheme  to  ``superposition of quasi-orthogonal-building-blocks" is straightforward.

\end{abstract}
\IEEEpeerreviewmaketitle
    \section{Introduction}
\label{sect-intro}
The Internet of Things (IoT)  is a novel paradigm  to 
support connections of billions of miscellaneous innovative devices and  will enable 
a range of new capabilities and services ranging from mission-critical services to massive
deployment of autonomous devices \cite{5billion}.
As an enabler of the IoT, 
machine type communication 
\cite{M2M-mag} which involves  a large number of user devices  that communicate autonomously with the aim of forming a ubiquitous and automatic communication network without human intervention, has received tremendous interest among mobile network operators,  equipment vendors, 
and research bodies.
In a M2M scenario,  
a cellular base-station (BS)  is required to connect to    a large number of devices  with typically sporadic  activity patterns \cite{M2M-mag}.
The sporadic traffic pattern may be due to  the fact that  devices are often designed to sleep most of the time in order to save energy and are activated only when triggered by external events,  as
is  typically the case in a sensor network.
Simultaneous random access attempts from massive sporadic machine-type communications (mMTC)  devices may severely congest a shared physical random access channel (PRACH)  in the current LTE and 5G networks \cite{OptimACB}
and  cause intolerable delay,  packet loss,  and even service unavailability. %

Various approaches have been proposed to    improve the performance of LTE networks serving MTC devices. Leya \emph{et al.}  proposed to split the random access pilots  into two sets to serve conventional data applications of user equipments (UEs)   and short data applications of the mMTC devices separately \cite{RA-app1}.\footnote{We utilize mMTC device and UE interchangeably throughout this work whenever there is no ambiguity.}
With the group paging approach,  Wei \emph{et al.}  introduced a model to estimate the number of successful and collided MTC devices in each random access slot  \cite{WeiRAC}.  Lee \emph{et al.} proposed a group-based RA scheme,  in which location-aware MTC devices located nearby are put into one group and tagged with the same group identifier \cite{GroupRAC}.  Group-based access methods can accommodate a large number of MTC devices   at the expense of prolonged access delay due to multi-hop transmissions. 
Slotted access proposed in \cite{SlotMAC} can mitigate congestion in light loads,  however,  it is not practical in massive connectivity load scenarios.
The 3rd Generation Partnership Project (3GPP)   introduced an access class barring (ACB)  scheme  \cite{ACB-3gpp} to reduce random access overload in cellular networks by broadcasting an ACB parameter $\rho$,   where $0\leq\rho\leq1$,   to all MTC devices via system information blocks (SIBs).  The load is relieved by permitting base stations  (BSs)  
to temporarily block low-priority devices,  or reduce their probability of  accessing the PRACH by a dynamic update of the relevant access parameters \cite{PRADA}.
Reference  \cite{CRMA-ambiguity}   proposed a  multi-preamble transmissions scheme with massive multiple-input multiple-output (MIMO)  to mitigate the pilot  collision at the cost of increased system resources and  codeword ambiguity  at receiver.
Massive MIMO opens up new avenues for enabling highly efficient random access
(RA)  by offering an abundance of spatial degrees of freedom \cite{Success-RA-MIMO}--
\cite{TwoStep-RA}.

Despite the approaches \cite{RA-app1}--\cite{TwoStep-RA} proposed from different angles, the size of  pilots   remains the   limiting factor that determines the  RA  success probability.
The conventional scheme dictates   that  the users contend for random access by uniformly choosing at random  one pilot from a pre-specified pool of  sequences  \cite{RAC-3GPP}.   
In 3GPP LTE systems, the pilots are constructed from prime-length Zadoff-Chu (ZC)  sequences with different cyclic shifts and root indexes \cite{PRACH}.

Allocating more time/frequency physical  resources  for RA purposes can alleviate the pilot collision, however,  it is obviously very costly in terms of spectral efficiency, which is of paramount importance
in  wireless system designs.
In this paper, we introduce  the  pattern-domain  pilot design  paradigm  {based on   a ``\emph{superposition of orthogonal-building-blocks}"}  to enlarge the size of contention space, without resorting to  increasing   physical resources.
Specifically, a pattern-domain pilot
is  designed based  on  the superposition of $K$  
 cyclically-shifted ZC sequences from  a root index.
The pattern-domain pilots are orthogonal between non-colliding patterns from the same root and exhibit  low  correlation values between patterns from different roots.
For a  cell with   $32$ cyclically-shifted ZC sequences and  $L =2$,   the contention  space is expanded by about $\binom{32}{2}/32= \frac{32!}{32*2! (32-2) !}\approx 16$  times. 
Since it is far less likely for multiple devices to choose an   identical pattern, i.e, \emph{L}  cyclically-shifted  ZC sequences,  than it is to choose a single ZC sequence (i.e., $L=1$ pattern), the pattern-domain based paradigm   can significantly reduce the pilot collision probability without sacrificing  \emph{excessively} on channel estimation (this will be elucidated later in Section \ref{sect-pdra-perf} and Section \ref{sect-Numerical}).
As a result, the overall performance of the mMTC RA   is enhanced. 

We present the framework of analysis of the  success probability of the proposed pattern division random access (PDRA)   scheme  in massive MIMO systems. Analysis illustrates that the proposed  PDRA   
  can reduce  the  pilot collision probability significantly  without compromising   excessively on the channel estimation and data detection performance.  Simulation    results demonstrate     that the PDRA  achieves an appreciable performance gain over the conventional one in massive
MIMO systems,  while keeping the existing  physical layer virtually intact.
  We can extend the ``{superposition of orthogonal-building-blocks}"  to  ``superposition of \emph{quasi-}orthogonal-building-blocks"
in an straightforward manner.

The rest of this work is organized as follows: Section \ref{sys-sec} describes the  system model. The  proposed  PDRA scheme is detailed in  Section \ref{pro-sect}. 
Numerical results  are provided in  Section \ref{sect-Numerical} and  are compared with the conventional scheme    in Section \ref{sect-Numerical}.
We conclude this work  in Section \ref{conclu-sec}.

\emph{Notation:} Vectors and matrices are denoted by boldface lowercase and uppercase letters,  respectively.
The transpose,  complex conjugate,  and conjugate transpose are represented by ${\left(\cdot  \right) ^{{T}}}$,   ${\left(\cdot  \right) ^*}$,   and ${\left(\cdot  \right) ^H}$,   respectively.
$\textbf{r}\sim \mathcal{CN}(\textbf{0},\textbf{R}) $  indicates that $\textbf{r}$ is a circularly symmetric complex Gaussian (CSCG)  random vector with zero-mean and covariance matrix \textbf{R}.  $\binom{m}{n}=\frac{m!}{n! (m-n) !}$ denotes the binomial coefficient.
$\left \lfloor{{x}}\right \rfloor$ represents the    greatest integer less than or equal to  \emph{x}.
   {
  $X \sim \Gamma(\alpha, \beta)  $ 
  denotes that  $X$  is gamma-distributed  with shape parameter  $\alpha$ and scale parameter
  $\beta$.} $\delta(\cdot) $ denotes  the Dirac delta function.

\section{System Model}
\label{sys-sec}
\begin{figure}[t]
\centering
\includegraphics[width=0.38\textwidth]{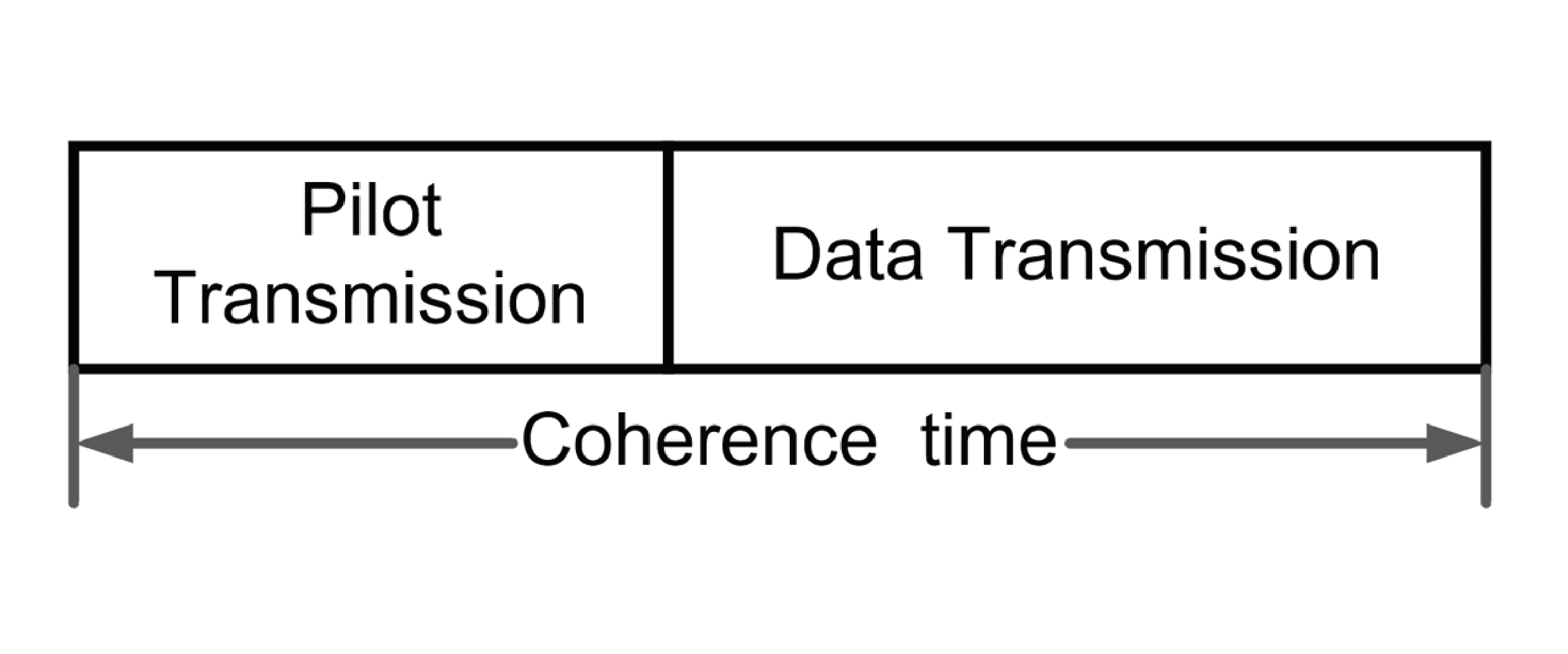} 
\label{qpsk-ber}
\centering
\vspace*{-5pt}
\caption{
Random access  transmission structure.
}
\label{fig-ber12}
\end{figure}

Consider a single-cell uplink M-MIMO system   where the BS  with $M\thinspace(\gg1) $ antennas serves
single-antenna mMTC devices.
The mMTC devices are assumed to be synchronized and each mMTC device decides in each coherence block 
whether or not to access the channel  with probability $\varepsilon$ in an independent and identically distributed (i.i.d.)   manner \cite{TwoStep-RA}.  
 On each RA occasion, we assume that there are  $N_{}$  UEs  active and  contend with their uplink payloads  by directly transmitting   pilots with length $N_{\rm{ZC}} $ randomly  selected from a pilot pool:  $\mathcal{S}=\left\{\textbf{s}_1,\textbf{s}_2,\cdots,\textbf{s}_{N_{\mathrm{P}}}\right\}$  
followed by  data transmission  
 as illustrated in Fig. \ref{fig-ber12}.  
The complex channel vector between the  BS  and the \emph{n}-th mMTC device
is given by ${\mathbf {f}_{n}}= \sqrt {\xi_n}  {{\mathbf {h}}}_{n} \in \mathbb {C}^{M}$,  where $\xi_n $  denotes the large-scale fading value 
and ${{\mathbf {h}}}_{n} \sim \mathcal {CN}(0,{\mathbf {I}}_{M}) $  
represents the small-scale fading coefficients. 
The channel is constant and frequency-flat for the transmission of pilot and data  samples. 

In commercial wireless systems,  base stations  transmit reference signals \cite{RAC-3GPP} that are used by the user equipment (UE)  to estimate the pathloss \cite{PC-3GPP}. 
The UE can then adjust its initial transmission power based on the measurement of the downlink reference signal power \cite{PC-3GPP}.
For the sake of analytical traceability,
we assume perfect power control 
of each active UE,\footnote{
    We abbreviate active UE as UE in later exposition if there is no ambiguity.}  the   {received}  pilot  $\mathbf {Y} \in \mathbb {C}^{M \times {N_{\mathrm{ZC}}}}$ is expressed as:
  \begin{equation}
   \mathbf {Y}=\sum _{n=1}^{N_{\mathrm{}}}\sqrt {P}\mathbf {h}_{n}\mathbf {s}^{\mathrm {T}}_{n}+\mathbf {W},
  \end{equation} 
where $P$ denotes the expected   {received}   power for all   UEs,
and  $\mathbf {W} \in \mathbb {C}^{M \times N_{\mathrm{ZC}}}$ represents the  noise matrix
with each  element  distributed as $\mathcal {CN}(0,\sigma ^{2}) $.
The uplink signal-to-noise ratio (SNR)  of each  UE is defined as   $\alpha_{\mathrm {R}}\triangleq P/\sigma ^{2}$.

The received data symbol vector $\mathbf {z} \in \mathbb {C}^{M}$ is given by:
\begin{equation}
  \label{y_rece} 
\mathbf {z} =\sum _{n=1}^{N_{}}\sqrt {P}\mathbf {h}_{n}{d}_{n}+\mathbf {n},
  \end{equation}
where  
  ${d}_{n}$ represents  the data symbol transmitted by the \emph{n}-th   UE and $\mathbb {E}[|d_{n}|^{2}]=1$,
  $\mathbf {n}$ denotes the background noise vector distributed as $\mathcal {CN}(0,\sigma ^{2}\mathbf {I}_{M}) $.

The BS is responsible for  detecting the active devices, estimating their channels, and decoding the transmitted data based on the acquired channel estimates to accomplish coherent data transmission. 
 If all the active UEs  select different pilots,
 the BS can detect the set of active UEs,  estimate their
channels and decode  data transmitted by the active UEs with adequate accuracy.
 Whereas, if multiple mMTC devices  select the same pilot,
the BS can not detect the collided  user channels
and the data detection will fail
at the BS  as a result of pilot collision.
The number of antennas \emph{M}  is assumed to   {be}    sufficiently large in this work, so mutual orthogonality among the vector-valued channels to the UEs can be approximately achieved. 

\section{Pattern Division Random Access (PDRA)   in  Massive MIMO Systems}
\label{pro-sect}
\subsection{Proposed Pattern-domain   Pilot Generation}
\label{sect-PRACH}
In 3GPP 4G LTE and 5G NR systems, the pilots are generated from  ZC sequences with different cyclic shifts and root indexes \cite{PRACH}. The set of pilots is first constructed from cyclic-shifts of a single root sequence, and then by cyclic shifts of different roots if necessary.
A ZC sequence $\textbf{c}_u $ of odd length $N_{\mathrm{ZC}}$  derived from root $u$ is defined as:
\begin{equation}
c_{u}[l]=e^{- {{j}}\frac {\pi u l(l+1)}{N_{\mathrm{ZC}}}}, ~0\leq l \leq N_{\mathrm{ZC}}-1,
 \end{equation}
where the root $u$  is relatively prime to $N_{\mathrm{ZC}}$. The  total number of roots is denoted by  ${{R}}$.

The  cyclic shift offset $N_{\mathrm{CS}}$ is dimensioned so  that the zero correlation zone (ZCZ)  of the sequences guarantees the orthogonality of the pilot sequences
regardless of the delay spread and time uncertainty of the UEs.
The   $N_{\mathrm{CS}}$ enables unambiguous detection with timing uncertainty $\tau<N_{\mathrm{CS}}$,  which is given by:
   \begin{equation}
  N_{\mathrm{CS}}\geq \left \lceil{\left ({\frac {2 R_c}{c}+\tau _{\max}}\right)  \times N_{\mathrm{ZC}} \times \Delta f_{\mathrm{RA}} +n_{\mathrm{g}}}\right \rceil,
\end{equation}
 where $R_c$ is the cell radius, $c\approx3\times10^8$ is the speed of light, $\tau_\mathrm{max}$ is the maximum expected delay spread of the propagation  channel, $\Delta f_{\mathrm{RA}}$ is the PRACH subcarrier spacing (SCS), and $n_g$ are additional guard samples due to the pulse shaping filter.
 The size of the pilot subset constructed from a single root ZC sequence $N_{\mathrm{SS}}$,  i.e., sequence subset size  is
$N_{\mathrm{SS}}=\left \lfloor{\frac {N_{\mathrm{ZC}}}{N_{\mathrm{CS}}}}\right \rfloor$.   

 ZC sequences  derived from a single root ZC sequence are generated as:
\begin{equation}
c_{u,v} [l]=c_{u} [(l+C_{v}) ~ \mathrm{mod}~ N_{\mathrm{ZC}} ],
\end{equation}
where $C_{v}=v N_{\mathrm{CS}}$ with $0\leq v \leq N_{\mathrm{SS}} -1$ represent the cyclic shifts.
We denote  ZC sequences   derived from root \emph{u} with cyclic shift \emph{v} as:
$\textbf{c}_{u,v}=\left\{c_{u,v} [0],\cdots,c_{u,v} [ N_{\mathrm{ZC}}-1]\right\}$.

The periodic autocorrelation of  the ZC sequence is zero for all off-peak shifts,  and the cyclically-shifted ZC sequences constructed by a root sequences are orthogonal. The cross-correlation value between  the ZC sequences   constructed from
different  root sequences is $\sqrt{N_{\mathrm{ZC}}}$ if $N_{\mathrm{ZC}}$ is a prime number.  

Increasing the  pilots allocated to the RA purpose  can   reduce the collision probability at the cost of resources for data transmission.
Accounting for this, we   expand  the contention  space    to the  pattern-domain  in a single random access sub-frame   without resorting to  increasing  the physical resources.
In this paper, we introduce the  pattern-domain  pilot design scheme  based on  a {``{superposition of orthogonal-building-blocks}"}   to enlarge the size of contention space. 
As an illustrative example,         we propose to utilize the ZC sequence with prime length as the \emph{building-block} to construct pattern with large pool size and
desirable   channel estimation property.
  Specifically, the pattern    derived   from root $u$ is
 constructed based on the superposition of the \emph{L} cyclically-shifted ZC sequences  as follows:
\begin{equation}
\label{equ-pd}
\begin{split}
{{\bf{s}}_{u,i}} &= \frac{1}{{\sqrt L}}\underbrace {\left({{{\bf{c}}_{u,v}} + {{\bf{c}}_{u,q}} +  \cdots {\rm{ +}}{{\bf{c}}_{u,s}}} \right)}_{\rm{Sum~of~\it{L}~terms}},
 v \ne q \ne  \cdots  \ne s \in \{0, \cdots,{N_{{\rm{SS}}}}{\rm{-}}1\},
0 < i < N_{\mathrm{PS}} - 1,
\end{split}
\end{equation}
where $N_{\mathrm{PS}}= \binom{{N_{{\rm{SS}}}}}{L} $ denotes the size of the pattern subset constructed from a single root, i.e., pattern subset size.
 The pilot pool $\mathcal{S}$  designed based on the  pattern-domain  paradigm can be expressed as follows:
\begin{equation}
\label{equ-pd-s}
\mathcal{S}=\left\{{{{\bf{s}}_{{u_1},0}} \cdots {{\bf{s}}_{{u_1},{N_{{\rm{PS}}}-1}}},{{\bf{s}}_{{u_2},0}} \cdots {{\bf{s}}_{{u_2},{N_{{\rm{PS}}}-1}}},{{\bf{s}}_{{u_{_{\rm{R}}}},0}} \cdots {{\bf{s}}_{{u_{_{\rm{R}}}},{N_{{\rm{PS}}}-1}}}} \right\},
\end{equation}
where $N_{\mathrm{P}}=RN_{{\rm{PS}}}$ represents the pattern-domain pilot pool size.
For the sake of illustration simplicity, we utilize ${\bf{s}}_{i}$  to denote ${\bf{s}}_{{u_r},i}$ with $ u_r \in\left\{u_1, u_2,\cdots,u_{\rm{R}}\right\}$  in later exposition if there is no ambiguity.
 Based on (\ref{equ-pd})  and (\ref{equ-pd-s}), the pattern-domain pilots  are orthogonal   between non-colliding patterns derived from the same root and exhibit low  correlation values between patterns derived from different roots.

 {The extension of the proposed ``superposition of the orthogonal-building-blocks" to    ``superposition of the   \emph{quasi-}orthogonal-building-blocks" is straightforward.
  We may extend the superposition operation to more generic finite-field operations which can be left for future investigation.
The developments in this work are first geared toward \emph{insights}
and then toward generality.
}

Compared to the conventional scheme, the   pilot pool size of the proposed pattern-domain  based paradigm  is  enlarged by:
\begin{equation}
\mu =\frac{{{R}\binom{{N_{{\rm{SS}}}}}{L}}}{{R}{N_{\rm{SS}}}}{ =}\frac{{\binom{{N_{{\rm{SS}}}}}{L}}}{{{N_{{\rm{SS}}}}}}.
\end{equation}

For the sake of simplicity, we only consider the pattern  based on $L =2$ cyclically-shifted ZC sequences  in this work.
For a typical cell configured with   $N_{\mathrm{SS}} =32$,   the size of the pilot pool $N_{\mathrm{P}}$ is   enlarged by about 16 times without allocating more time/frequency physical  resources.

 Increasing the size of the pilot pool is not a complete solution to the problem of the RA  success in a crowded cellular network since the RA performance depends not only on the
probability of pilot collision 
but also on the probability of successful  data detection.
We elucidate  the underlying idea of   the proposed PDRA  with  massive MIMO systems in Section \ref{sect-pdra-perf}.

\subsection{Success Probability of PDRA   in   Massive MIMO Systems}
\label{sect-pdra-perf}

For the sake of simplicity,  the matched filter (MF)  detector  is applied  at the BS to  perform channel estimation and data detection.
In order to successfully decode the data of an active UE, 
 it is of paramount importance for the BS to  acquire accurate channel state information (CSI)  of  active UEs in  the case of no  pattern collision and then performs  data detection with adequate accuracy to ensure active UEs  exhibit sufficiently high signal-to-interference-and-noise-ratios (SINRs).  
The success probability  of the  RA depend on the following two factors: 
\begin{enumerate}
  \item Pattern collision:
For a  PDRA scheme,   only if all  ZC  sequence components   of an active   UE collide with other  UEs',  the active  UE  will have no chance of recovering the data. In other words, the active UE can still  obtain reasonably accurate CSI even with  only one collision-free ZC  sequence component.
  \item Adequate SINR:  Data transmission of an active   UE is considered successful only when
its     SINR is sufficiently high. 
\end{enumerate}

Based on the aforementioned analysis, we define the  probability of no pattern collision $P_{\rm{S}}$ and $P\left( {\alpha _{{\rm{MF}}}  \geq  \alpha _{{\rm{Th}}}^{}} \right) $ for 
the active UE  as the success probability, where $\alpha _{{\rm{MF}}}$ is the    SINR  of  the  active  UE and $\alpha_{\mathrm{Th}}$ is a given SINR threshold which depends  on the modulation and
coding schemes of the specific UE.  {(The subscript MF  indicates that the matched filter  is used as the detector  at the base station.)}

Without loss of generality,  the  success probability of the first   UE  ${P_{{\rm{MF}}}}$ is derived as follows:
\begin{align}
 \label{ZF-detec-PDRA}
{P_{{\rm{MF}}}} = {P_{\rm{S}}}P\left({\alpha _{{\rm{MF}}}^1\geq    {\alpha _{{\rm{Th}}}}} \right)
   =   {P_{\rm{S}}} \sum\limits_{K = 0}^{N - 1} { P\left( K  \right)}  \left[  {{P_{{E_0}}}P \left(  {\alpha _{{\rm{MF}}}^1  \geq  {\alpha _{{\rm{Th}}}}\left| {K,{E_0}} \right.}  \right)   +  {P_{{{{E}}_{{1}}}}} P \left({\alpha _{{\rm{MF}}}^1  \geq  {\alpha _{{\rm{Th}}}} \left| {K,{E_1}} \right.} \right)}  \right],
\end{align}

\begin{itemize}

   \item where
$ {{P}}_{\mathrm {S}}$ denotes the probability that no pattern collision occurs between  the first  UE and the other $N_{} -1$  UEs 
and is  expressed as: 
\begin{equation}
\label{pro-suc}
 {{P}}_{\mathrm {S}}=\left ({1-\frac {1}{RN_{\mathrm{PS}}}}\right) ^{N_{}-1}=\left ({1-\frac {1}{N_{\mathrm{P}}}}\right) ^{N_{}-1},
 \end{equation}
  \item  ${{P}}(K) $ denotes the probability that $K$ other   UEs choose the other ${R}-1$ different root patterns 
from the first  UE and the rest $N_\mathrm{}-1-K$  UEs select the same root pattern  
 as the first active UE  and is given by:
\begin{align}
\label{pro-K}
{{P}}(K) =\frac {{\binom {N_\mathrm{}-1}{K}}(({R}-1) N_{\mathrm{PS}}) ^{K}(N_{\mathrm{PS}}-1) ^{N_\mathrm{}-1-K}}{({R}N_{\mathrm{PS}}-1) ^{N_\mathrm{}-1}},
 \end{align}
       \item  $E_0$  denotes the event    whereby the $N-1-K$ users  have  no      single ZC sequence component  colliding  with the first UE and  $P_{E_0}$ represents the probability of the event $E_0$ occurring, which is expressed as:
             \begin{equation}
  \label{p0}
    {P_{{E_0}}} = \frac{{{{\left({({N_{{\rm{SS}}}} - {\rm{2}}) ({N_{{\rm{SS}}}} - 3) /2} \right)}^{N - 1 - K}}}}{{{{\left({{N_{{\rm{PS}}}} - {\rm{1}}} \right)}^{N - 1 - K}}}},
  \end{equation}

   \item $ {P}(\alpha ^{1}_{\mathrm {MF}} \geq \alpha _{\mathrm {Th}}|K,E_0) $  
  represents  the probability of $\alpha ^{1}_{\mathrm {MF}} \geq \alpha _{\mathrm {Th}}$ on the condition that \emph{K} other  UEs choose the other ${R}-1$  different root patterns from the first active  UE and the event $E_0$ occurs.
\normalsize
    \item $E_1$    denotes the event    whereby  the $N-1-K$ users  have  one      single ZC sequence component  colliding  with the first UE and  $P_{E_1}$ represents the probability of the event $E_1$ occurring, which is given by:
\begin{equation}
\label{pr1}
{\begin{split}
&{P_{{\rm{}}E_1}}=
\frac{{2   \sum\limits_{T = 1}^{N - 1 - K}
{
\binom{N - 1 - K}{T}
{{({N_{{\rm{SS}}}}  -  {\rm{2}})}^T}{{\left({\left({{N_{{\rm{SS}}}}  -  2} \right) \left({{N_{{\rm{SS}}}}  -  3} \right) /2} \right)}^{N - 1 - K - T}}}}}{{{{\left({{N_{{\rm{PS}}}} - {\rm{1}}} \right)}^{N - 1 - K}}}},
\end{split}
}
\end{equation}

\normalsize

  \item  $ {P}(\alpha ^{1}_{\mathrm {MF}} \geq \alpha _{\mathrm {Th}}|K,E_1) $  
  represents  the probability of $\alpha ^{1}_{\mathrm {MF}} \geq \alpha _{\mathrm {Th}}$ on the condition  that \emph{K} other  UEs choose the other ${R}-1$  different root patterns from the first active  UEand the event $E_1$ occurs. \newline  \indent The detailed derivation of  ${P_{{E_0}}}$ and  ${P_{{E_1}}}$ is deferred  to the Appendix Section.
  \end{itemize}

   For the sake of illustration clarity, we denote $\textbf{s}_{k}$  as  the pattern used by UE \emph{k} with  a   slight abuse of notation. Without loss of generality, we assume that the first   UE utilizes pattern $\textbf{s}_{1}  (=\frac{1}{\sqrt{2}}(\textbf{c}_{1,1}+\textbf{c}_{1,2})) $.

    In the case of event $E_0$ occurring,  multiplying $\bf{Y}$  from the right-side by    $\frac{{{\bf{s}}_1^*}}{{\left\| {{{\bf{s}}_1}} \right\|}}$,  we obtain the    estimated channel  vector of the active UE1 as follows: 

\begin{equation}
  \begin{split}
{{\bf{g}}_{1,{E_0}}}
&= {\bf{Y}}\frac{{{\bf{s}}_1^*}}{{\left\| {{{\bf{s}}_1}} \right\|}}
=\sqrt {{P_{}}} {{\bf{h}}_{\rm{1}}}{\bf{s}}_1^T\frac{{{\bf{s}}_1^*}}{{\left\| {{{\bf{s}}_1}} \right\|}}{ +}\underbrace {\sum\limits_{k = 2}^{K + 1} {\sqrt {{P}} {{\bf{h}}_k}{\bf{s}}_k^T\frac{{{\bf{s}}_1^*}}{{\left\| {{{\bf{s}}_1}} \right\|}}}}_{K{\rm{~ UEs~ with~  different ~ roots}}}
+  \underbrace {\sum\limits_{m = K + 2}^N   {\sqrt {{P_{}}} {{\bf{h}}_m}{\bf{s}}_m^T\frac{{{\bf{s}}_1^*}}{{\left\| {{{\bf{s}}_1}} \right\|}}}}_{N - 1 - K {\rm{~ UEs~  with ~ the}} {~ \rm{same}}~  {\rm{root}}}     + {\bf{W}}\frac{{{\bf{s}}_{1}^*}}{{\left\| {{{\bf{s}}_{1}}} \right\|}}\\
&=\sqrt {{P_{}}} {{\bf{h}}_{\rm{1}}}{\bf{s}}_1^T\frac{{{\bf{s}}_1^*}}{{\left\| {{{\bf{s}}_1}} \right\|}}{\rm{ +}}  \underbrace {\sum\limits_{k = 2}^{K + 1} {\sqrt {{P_{}}} {{\bf{h}}_k}{\bf{s}}_k^T\frac{{{\bf{s}}_1^*}}{{\left\| {{{\bf{s}}_1}} \right\|}}}}_{K{\rm{~ UEs~ with~  different ~ roots}}}   + {\bf{W}}\frac{{{\bf{s}}_1^*}}{{\left\| {{\bf{s}}_1^*} \right\|}}
=\sqrt {{P_{}}{N_{{\rm{ZC}}}}} {{\bf{h}}_1} + {\sum\limits_{k = 2}^{K + 1} {2\sqrt {{P_{}}} {{\bf{h}}_k}}} + {\bf{\tilde w}},
\end{split}
  \label{equ-e0}
 \end{equation}
   where $ \boldsymbol{\tilde{\mathrm{w}}} =\mathbf {W}\frac {{\textbf{s}}_{1}^{*}}{\|  {\textbf{s}}_{1} \|}   \sim \mathcal {CN}(\mathbf {0},   \sigma^2\mathbf {I}_{M}) $.

With the MF  receiver, the estimate of the data symbol of the first UE  
 is given by:
   \begin{equation}
   \label{x-est-equ}
   {d_1} = {{\bf{g}}^H_{1,{E_0}}}{\bf{z}} = \sqrt {{P}} {{\bf{g}}^H_{1,{E_0}}}{{\bf{h}}_1}{d_1} + \sum\limits_{n = 2}^N {\sqrt {{P}} {{\bf{g}}^H_{1,{E_0}}}{{\bf{h}}_n}{d_n} + {{\bf{g}}^H_{1,{E_0}}}{\bf{n}}}.
 \end{equation}
    Utilizing  (\ref{x-est-equ}), we obtain   the    SINR of the  first UE $ \alpha _{{\rm{MF}}}^1$ as:
 \begin{equation} 
\label{MF-est-equ}
 \alpha _{{\rm{MF}}}^1=\frac{{{P}{{\left| {{{\bf{g}}^H_{1,{E_0}}}{{\bf{h}}_1}} \right|}^{\rm{2}}}}}{{\sum\limits_{n = 2}^N {{P}{{\left| {{{\bf{g}}^H_{1,{E_0}}}{{\bf{h}}_n}} \right|}^{\rm{2}}} + {{\left\| {{{\bf{g}}^H_{1,{E_0}}}} \right\|}^2}{\sigma ^{\rm{2}}}}}}
 \stackrel {M \to \infty} =  \frac{{{N_{{\rm{ZC}}}}}}{{4K}}.
 \end{equation}
  Equation (\ref{MF-est-equ})  is derived   based on the fact $\frac {\mathbf {h}_{n}^{\mathrm {H}}\mathbf {h}_{n}}{M} \stackrel {M \to \infty}{\longrightarrow} 1$ and $\frac {\mathbf {h}_{n}^{\mathrm {H}}\mathbf {h}_{k}}{M} \stackrel {M \to \infty}{\longrightarrow} 0$ when $n\neq k$.
 The detailed derivation of  $ \alpha _{{\rm{MF}}}^1$ is deferred  to the Appendix Section.
Based on (\ref{MF-est-equ}),  the asymptotic value of  $P\left({\alpha _{{\rm{MF}}}^1\geq    {\alpha _{{\rm{Th}}}}\left| {K,{E_0}} \right.} \right) $ can be expressed as:
 \begin{equation}
 \label{MF-est-equ1}
\mathop {\lim}\limits_{M \to \infty}P\left({\alpha _{{\rm{MF}}}^1\geq    {\alpha _{{\rm{Th}}}}\left| {K,{E_0}} \right.} \right) ={1_{K ~\le \frac{N_{\rm{ZC}}}{{{4\alpha _{{\rm{Th}}}}}}}} = \left\{\begin{array}{l}
1,{\rm{~if~ \emph{K}}} \le \frac{N_{\rm{ZC}}}{{{4\alpha _{{\rm{Th}}}}}},\\
0,{\rm{~otherwise}}{.}
\end{array} \right.
 \end{equation}

 In the case of event $E_1$ occurring,  
  multiplying $\bf{Y}$  from the right-side by  $\frac{{{\bf{c}}_{1,1}^*}}{{\left\| {{{\bf{c}}_{1,1}}} \right\|}}$,    we obtain the    channel  vector of the first UE as follows:
    \begin{equation}
            {\begin{split}
{{\bf{g}}_{1,{E_1}}} & =  {\bf{Y}}\frac{{{\bf{c}}_{1,1}^*}}{{\left\|
{{{\bf{c}}_{1,1}}}  \right\|}}
{\rm{= }}\sqrt {{ \frac{P}{2}}} {{\bf{h}}_{\rm{1}}} ({\bf{c}}_{1,1} + {\bf{c}}_{1,2}) ^T \frac{{{\bf{c}}_{1,1}^*}}{{\left\|  {{{\bf{c}}_{1,1}}}  \right\|}}
{\rm{+}}\underbrace {\sum\limits_{k = 2}^{K  +  1}   {\sqrt {{ \frac{P}{2}}} {{\bf{h}}_k}({\bf{c}}_{k,1} + {\bf{c}}_{k,2}) ^T \frac{{{\bf{c}}_{1,1}^*}}{{\left\| {{{\bf{c}}_{1,1}}} \right\|}}}}_{K{\rm{~ UEs~ with~  different ~ roots}}}\\
  &+    \underbrace {\sum\limits_{m = K + 2}^N   {\sqrt {{\frac{P}{2}}} {{\bf{h}}_m}({\bf{c}}_{m,1}+{\bf{c}}_{m,2}) ^T\frac{{{\bf{c}}_{1,1}^*}}{{\left\| {{{\bf{c}}_{1,1}}} \right\|}}}}_{\tiny{N - 1 - K {\rm{~ UEs~  with ~ the}} {~ \rm{same}}~  {\rm{root}}}} + {\bf{W}}\frac{{{\bf{c}}_{1,1}^*}}{{\left\| {{{\bf{c}}_{1,1}}} \right\|}}
\\&=\sqrt {P\frac{N_{\rm{ZC}}}{2}} {{\bf{h}}_1} + {\sum\limits_{k = 2}^{K + 1} 2{\sqrt {{\frac{P}{2}}} {{\bf{h}}_k}}} + {\bf{\bar{w}}},
\end{split}
}
  \label{equ-e1}
 \end{equation}
   where $ {\bf{\bar{w}}}={\bf{W}}\frac{{{\bf{c}}_{1,1}^*}}{{\left\| {{{\bf{c}}_{1,1}}} \right\|}}  \sim \mathcal {CN}(\mathbf {0},   \sigma^2\mathbf {I}_{M}) $.
\noindent Similar to that of (\ref {MF-est-equ}), we can obtain the same   SINR value  for the event $E_1$.

Substituting (\ref{pro-suc})--(\ref{pr1}), (\ref{MF-est-equ})   and (\ref{MF-est-equ1})   into (\ref{ZF-detec-PDRA}), we obtain  the
success probability of the  ${P_{{\rm{MF}}}}$  as follows: 
{  \begin{align}   
   \label{MF-asym}
{P_{{\rm{MF}}}}& = {P_{\rm{S}}} \sum\limits_{K = 0}^{N - 1}  {P(K) \left[ {{P_{E_0}}P \left( {\frac{{{N_{{\rm{ZC}}}}}}{{4K}}  \ge  {\alpha _{{\rm{Th}}}}} \right)   +  {P_{E_1}}P \left( {\frac{{{N_{{\rm{ZC}}}}}}{{4K}} \ge
 {\alpha _{{\rm{Th}}}}} \right)} \right]}    =  {P_{\rm{S}}} \sum\limits_{K = 0}^{N - 1} {\left[ {P(K) \left( {{P_{E_0}}  +  {P_{E_1}}} \right) P \left({\frac{N_{{\rm{ZC}}}}{{4K}} \ge {\alpha _{{\rm{Th}}}}} \right)} \right]}  \nonumber\\
 & =  {P_{\rm{S}}}\sum\limits_{K = 0}^{N - 1} {\left[ {P(K) \left({{P_{E_0}} + {P_{E_1}}} \right) {1_{K \leq \frac{{{N_{{\rm{ZC}}}}}}{{4{\alpha _{{\rm{Th}}}}}}}}} \right]} = {P_{\rm{S}}}\sum\limits_{K = 0}^{\min \left\{ {\left\lfloor {\frac{{{N_{{\rm{ZC}}}}}}{{4{\alpha _{{\rm{Th}}}}}}} \right\rfloor,N - 1} \right\}} {P(K) \left( {{P_{E_0}} + {P_{E_1}}} \right)} \nonumber\\
 &  =  {\left({1 - \frac{1}{{{N_P}}}} \right) ^{N - 1}}\sum\limits_{K = 0}^{\min \left\{{\left\lfloor {\frac{{{N_{\rm{ZC}}}}}{{4{\alpha _{{\rm{Th}}}}}}} \right\rfloor,N - 1} \right\}} {\frac{{
\binom{N - 1}{K}
{{\left({(R - 1) {N_{{\rm{PS}}}}} \right)}^K}{{\left({{N_{{\rm{PS}}}} - {\rm{1}}} \right)}^{N - 1 - K}}}}{{{{\left( {R{N_{{\rm{PS}}}} - {\rm{1}}} \right)}^{N - 1}}}}} \nonumber\\
& \times \left(\frac{{{{\left({({N_{{\rm{SS}}}} - {\rm{2}}) ({N_{{\rm{SS}}}} - 3) /2} \right)}^{N - 1 - K}}}}{{{{\left( {{N_{{\rm{PS}}}} - {\rm{1}}} \right)}^{N - 1 - K}}}}+\frac{{2
 \sum\limits_{T = 1}^{N - 1 - K} {
   \binom{N - 1 - K}{T}
{{({N_{{\rm{SS}}}} - {\rm{2}})}^T}{{\left({\left({{N_{{\rm{SS}}}} - 2} \right) \left({{N_{{\rm{SS}}}} - 3} \right) /2} \right)}^{N - 1 - K - T}}}}}{{{{\left({{N_{{\rm{PS}}}} - {\rm{1}}} \right)}^{N - 1 - K}}}}
\right).
\end{align}
}
We will verify the accuracy of the asymptotic analysis of (\ref{MF-asym})  in  Section \ref{sect-Numerical} and study its behaviour for a finite (but large)  \emph{M}.

\section{Numerical Results}
\label{sect-Numerical}

In this section, numerical results are first carried out to verify the
asymptotic analysis of the   PDRA  in Section \ref{sect-pdra-perf}.
  We then carry out  system simulation  
of  
 the success probability  PDRA 
  and  compare it  with the  conventional one with various   parameters.   \subsection{Asymptotic Analysis  of PDRA} 
\begin{figure}[t]
\begin{narrow}{-0.00in}{0in}
\centering
\includegraphics[width=0.529\textwidth]{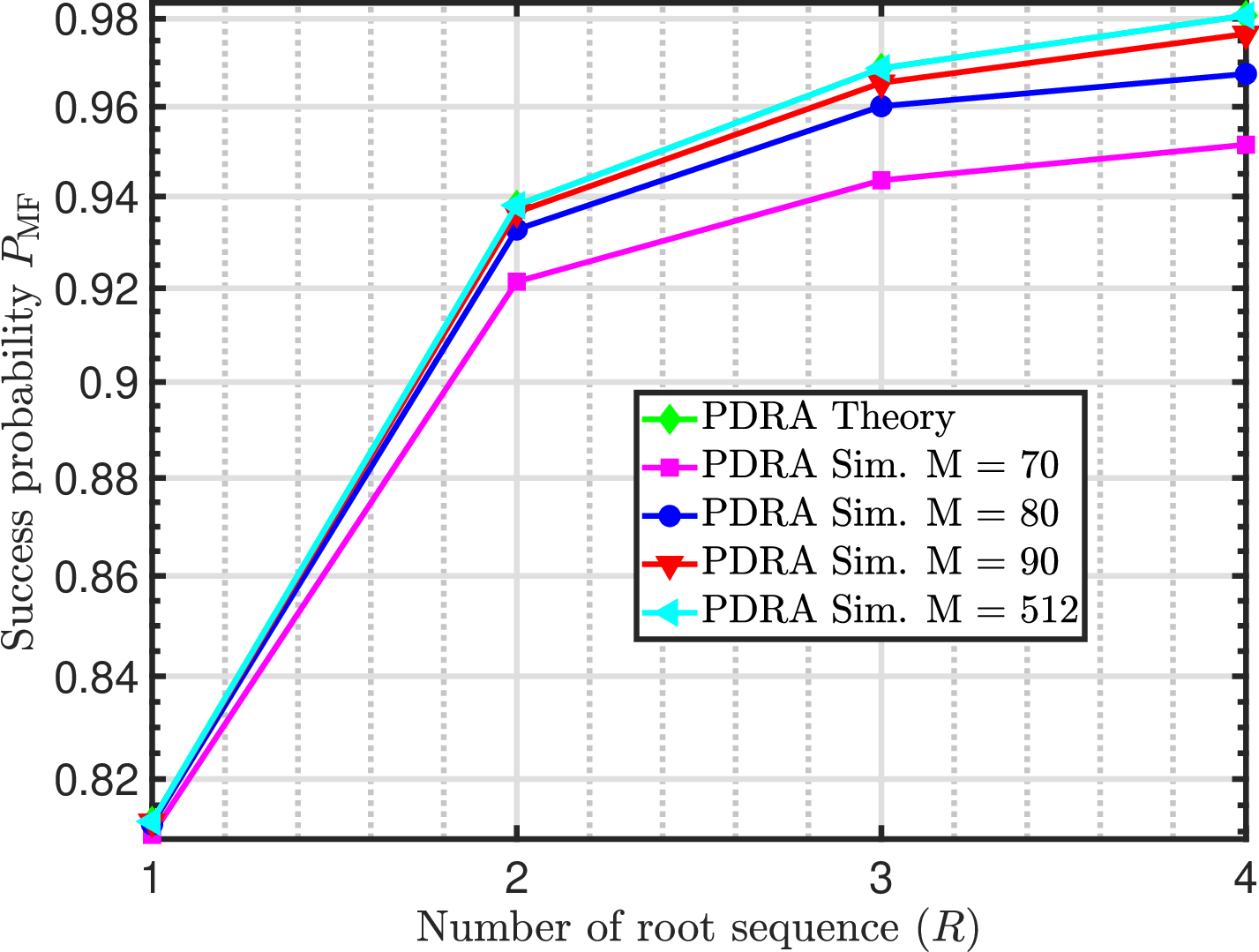} 
\label{qpsk-ber}
\caption{$P_{\rm{MF}}$  as a function of $R$ with different values of $M$, 
 when   $N_{\rm{SS}} =32,   N=10,$ and $ \alpha_{\rm{Th}}= 5\thinspace \rm{dB}$.
}
\label{fig-2}
\end{narrow}
\end{figure}

 Numerical results are first carried out to verify the
asymptotic analysis of the   PDRA  in Section \ref{sect-pdra-perf}. 
Fig. \ref{fig-2} illustrates the success probabilities as a function of \emph{R} with different values
of $M$ when $ N_{\rm{SS}} =32,  N=10,$ and $ \alpha_{\rm{Th}}= 5\thinspace \rm{dB}$. 
Fig. \ref{fig-2} shows that the numerical results (abbreviated as Sim.)  
closely match the analysis of  {(\ref{MF-asym})}  (labelled  as Theory)  of ${P_{{\rm{MF}}}}$  in Section \ref{sect-pdra-perf} for large \emph{M} and tighter results are observed  as \emph{M} grows. {The result of $M=512 $ antennas is almost indistinguishable from that of the theoretical analysis [c.f. {(\ref{MF-asym})}].}    As expected, the gap between the  numerical and analytical results slightly increases as the \emph{M} decreases since the mutual orthogonality among the vector-valued channels are not fully satisfied in that case.


\begin{figure}[t]
\begin{narrow}{-0.00in}{0in}
\centering
\includegraphics[width=0.255\textwidth]{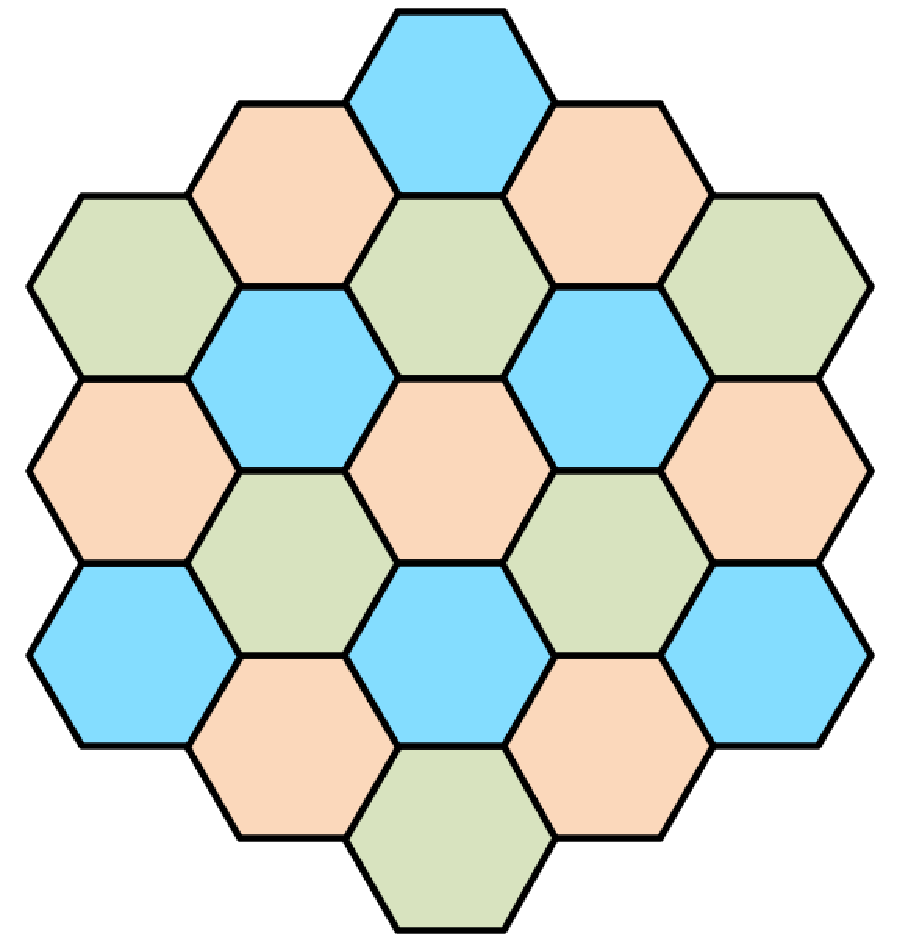} 
\label{qpsk-ber}
\caption{Illustration of the cell topology.
}
\label{fig-22}
\end{narrow}
\end{figure}
\subsection{System Simulations  Parameters}
\label{sect-NumS}
 {
 In this section, we show numerically how the proposed PDRA  performs in cellular networks.
We consider the scenario which is composed of 19 hexagonal grid shaped cells, distributed in
three different tiers as depicted in Fig. \ref{fig-22}.
 Two different channel models are considered in the simulations.
  The first one is uncorrelated Rayleigh fading, where $\mathbf {f}_{n}  \sim \mathcal {CN}(0,{\mathbf {I}}_{M}) $.
 The second one is the spatially correlated Rayleigh fading channel with $\mathbf {f}_{n} \sim \mathcal {CN}(\mathbf {0},\xi_{n} \mathbf {R}_{n}) $,   where we consider
a uniform linear array (ULA)  at the BS modeled by the exponential correlation model
with the correlation  coefficient $\rho$ between adjacent antennas
$\left [{\mathbf {R}_{n}}\right ]_{i,j} = \rho^{- | j-i |} e^{\jmath \delta _{n} (j-i)}$,
where $\delta_{n}$ is the angle between  BS and active UE \emph{n}. 
 This represents a non-line-of-sight (NLoS)  scenario with spatial correlation, indicating that the channel is statistically stronger in some spatial directions (determined by $\delta_n$)  than other directions.
The pathloss is modeled based on the urban micro scenario  \cite{Spatial Channel Model-3gpp}. The Rayleigh fading cases have a pathloss exponent of 3.8
and shadow fading with log-normal distribution and a standard
deviation of 10 dB, while the line-of-sight (LoS)  case has a pathloss  exponent of 2.5
and log-normal variations with a standard deviation of 4$\thinspace$dB.
The radius of each hexagon is $500\thinspace{\rm{m}}$,  and the UEs are uniformly distributed in the  cell at locations further than 30$\thinspace$m from the BS. 
We consider a scenario with  $ 10,000$
inactive UEs in the cell, where each UE accesses the network
with   
probability  $P_A$   in a given RA opportunity.
The signal-to-noise-ratio (SNR)  is defined as the expected pattern to noise power ratio at each antenna port of the BS.}
The  pattern length $N_{\rm{ZC}} $ is set to be 839 as the existing RA in 4G LTE and 5G NR standards \cite{PRACH}.

\begin{figure}[!t]
\begin{narrow}{-0.00in}{0in}
\centering
\includegraphics[width=0.529\textwidth]{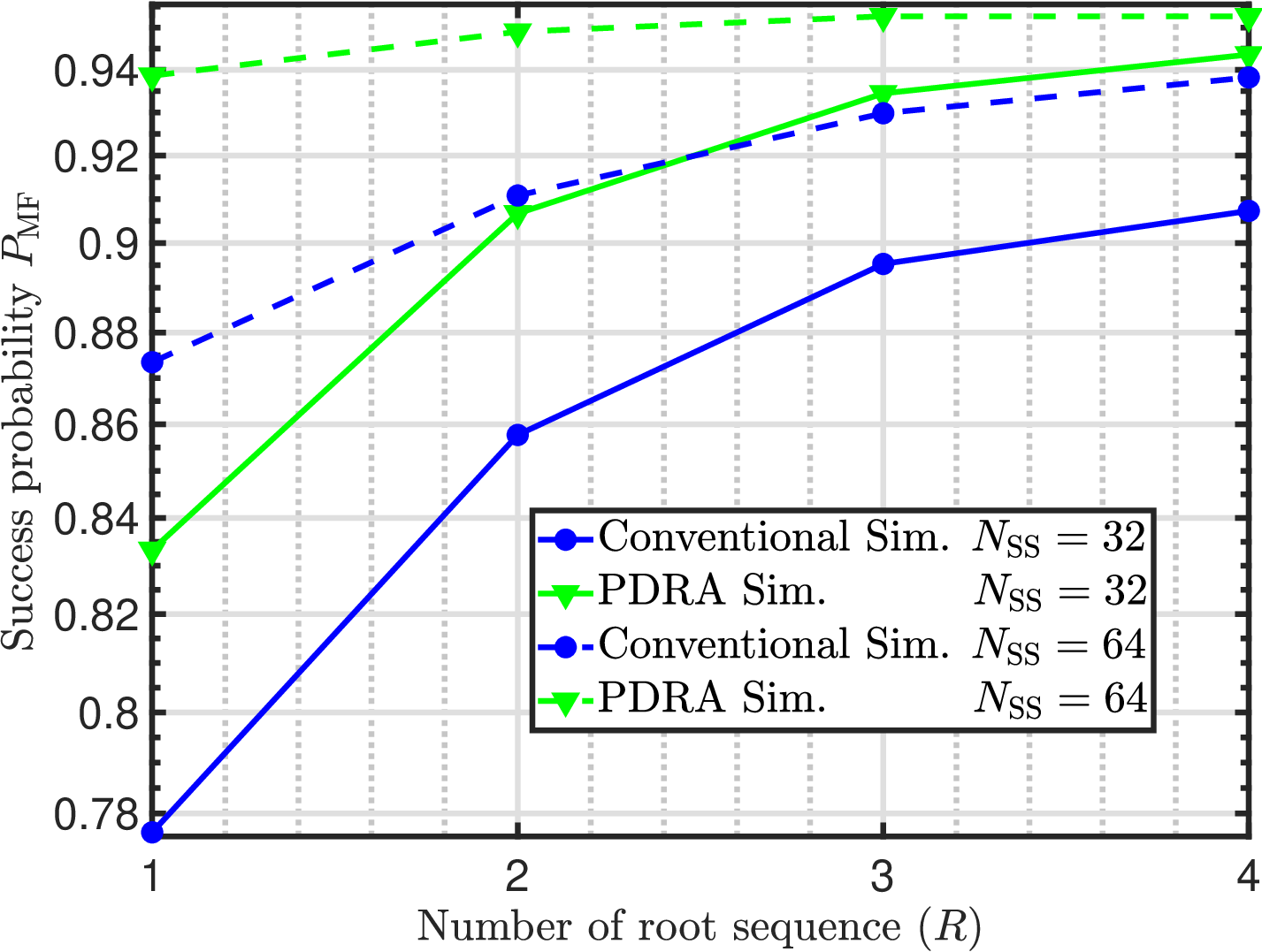} 
\label{qpsk-ber}
\caption{$P_{\rm{MF}}$  as a function of root number $R$  with different values of  $N_{\rm{SS}}$,   when   $ M =128,$  $ \alpha_{\rm{Th}}= 5\thinspace \rm{dB}$ and $ P_A=0.1\%$.
}
\label{fig-3}
\end{narrow}
\end{figure}

\begin{figure}[!t]
\begin{narrow}{-0.00in}{0in}
\centering
\includegraphics[width=0.59\textwidth]{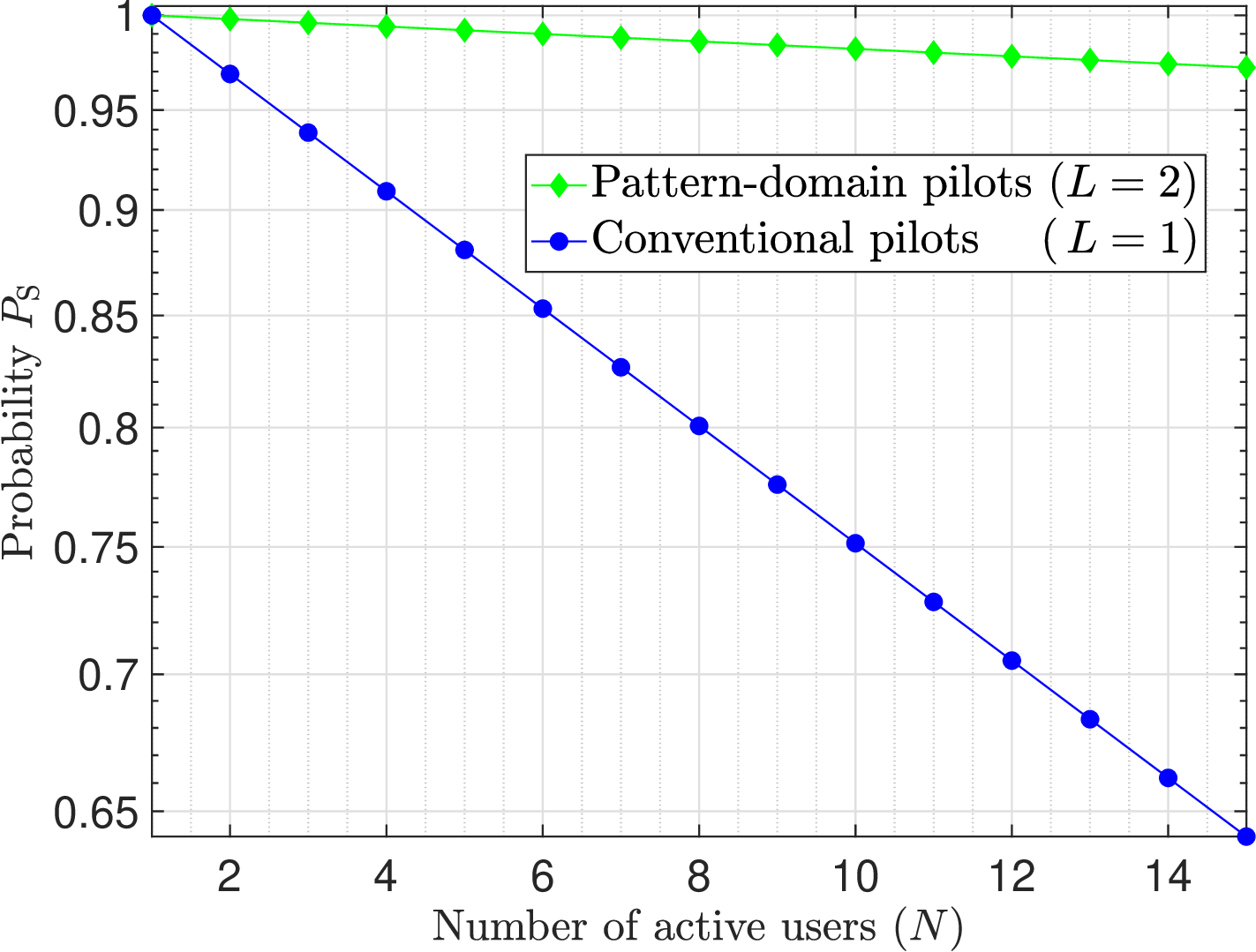} 
\label{qpsk-ber}
\caption{
$P_{\rm{S}}$  as a function of  of active users with different values of \emph{L}, when $N_{\rm{SS}} =32$.
}
\label{fig-4}
\end{narrow}
\end{figure}

\begin{figure}[!t]
\begin{narrow}{0.00in}{0in}
\centering
\includegraphics[width=0.549\textwidth]{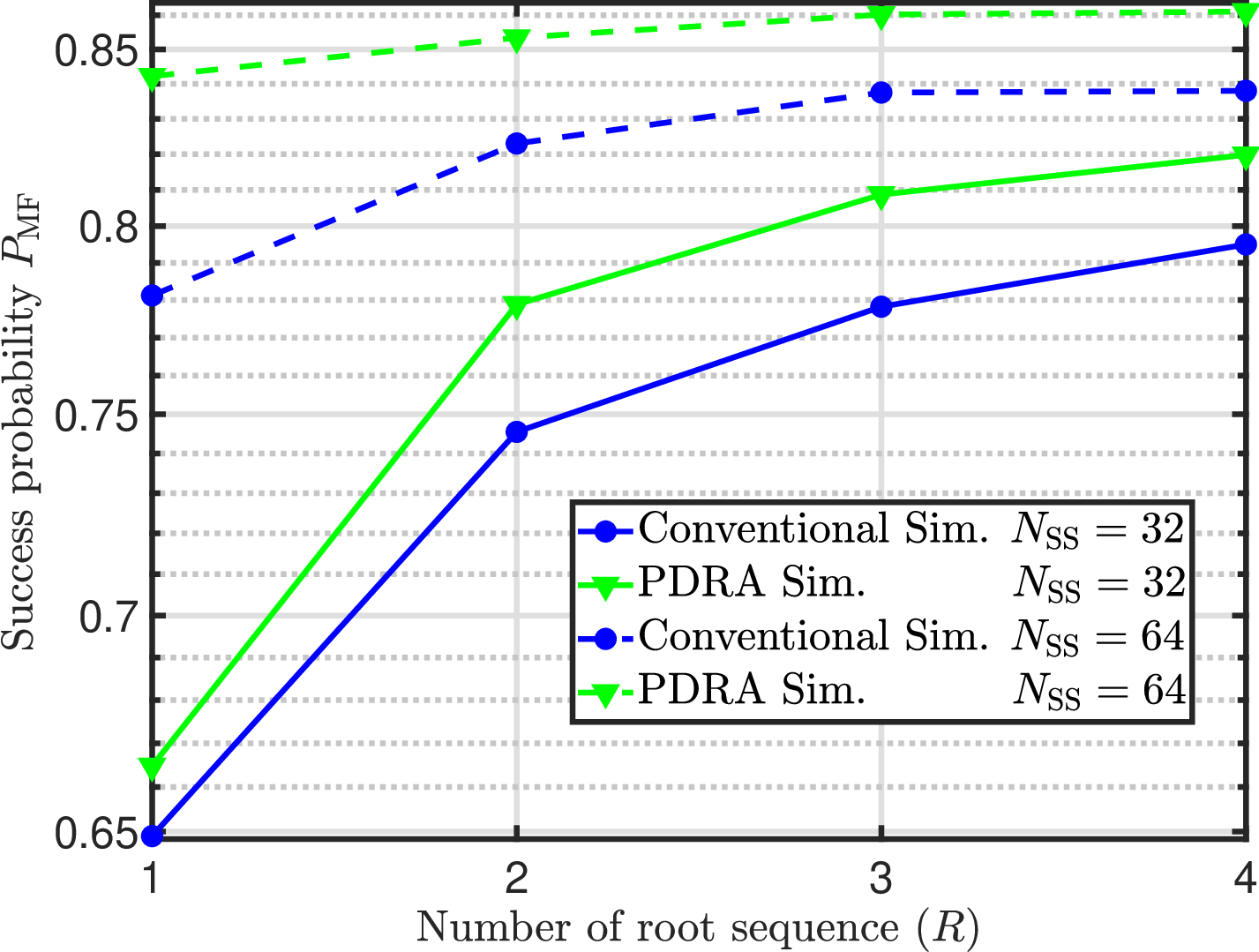} 
\label{qpsk-ber}
\caption{
$P_{\rm{MF}}$  as a function of root number $R$  with different values of  $N_{\rm{SS}}$,     when   $ M =128,$  $ \alpha_{\rm{Th}}= 5\thinspace \rm{dB}$ and $ P_A=0.15\%$.
}
\label{fig-5}
\end{narrow}
\end{figure}

\subsubsection{Success Probability  Under Uncorrelated Rayleigh Fading Channel}
Fig. \ref{fig-3} illustrates  the success probabilities as a function of   $R$ with different values of $N_{\rm{SS}}$. 
As the $N_{\rm{SS}}$ increases, the success probability of both schemes improves
and 
the proposed PDRA achieves an appreciable performance gain over the  conventional one as illustrated  in Fig. \ref{fig-3}.   It is also shown in  Fig. \ref{fig-3} that the PDRA with   $N_{\rm{SS}} =32$ {even slightly}  outperforms the conventional one with  $N_{\rm{SS}} =64$ for  $R>2$.

This can be explained as follows:  The success probability of ${P_{{\rm{MF}}}}$ is determined by two factors: 1)  $ {{P}}_{\mathrm {S}} $; and 2)  $P\left({\alpha _{{\rm{MF}}}^1 \geq  \alpha _{{\rm{Th}}}^{}} \right) $.
  Increasing the number of pilots  can alleviate the pattern collision $ {{P}}_{\mathrm {S}} $ as shown in Fig. \ref{fig-4}, however, the  associated non-orthogonality also degrades the channel estimation and data detection performance [c.f. (\ref{MF-est-equ})].
  Nonetheless, the performance gain derived  from the  collision probability reduction (due  to the  increase of contention space)   outweighs  
  the degradation of  channel estimation  and data detection caused by the  non-orthogonality of the patterns. 

  A similar phenomenon  can be observed  for a higher user activity probability $P_A$ 
  as illustrated in Fig. \ref{fig-5}.

\begin{figure}[t]
\begin{narrow}{-0.55in}{0in}
\subfigure[$\rho=0.7,N_{\rm{SS}}=32$]{\begin{minipage}[t]{0.5615\textwidth}
\centering
\includegraphics[width=1\textwidth]{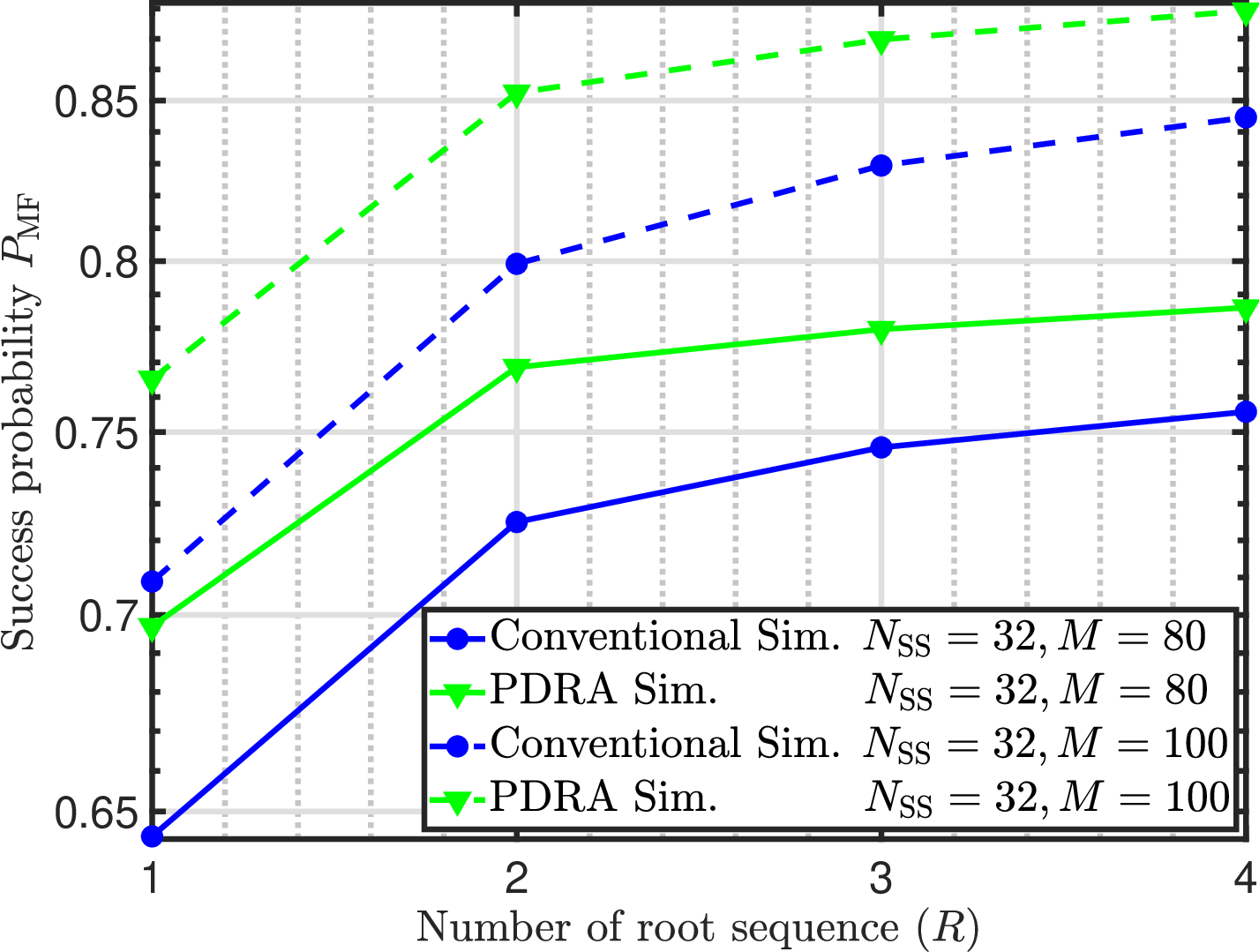} 
\label{qpsk-ber}
\label{fig-60}
\end{minipage}}\hspace*{12pt}
\subfigure[$\rho=0.7, N_{\rm{SS}}=64$]{\begin{minipage}[t]{0.5615\textwidth}
\centering
\includegraphics[width=1\textwidth]{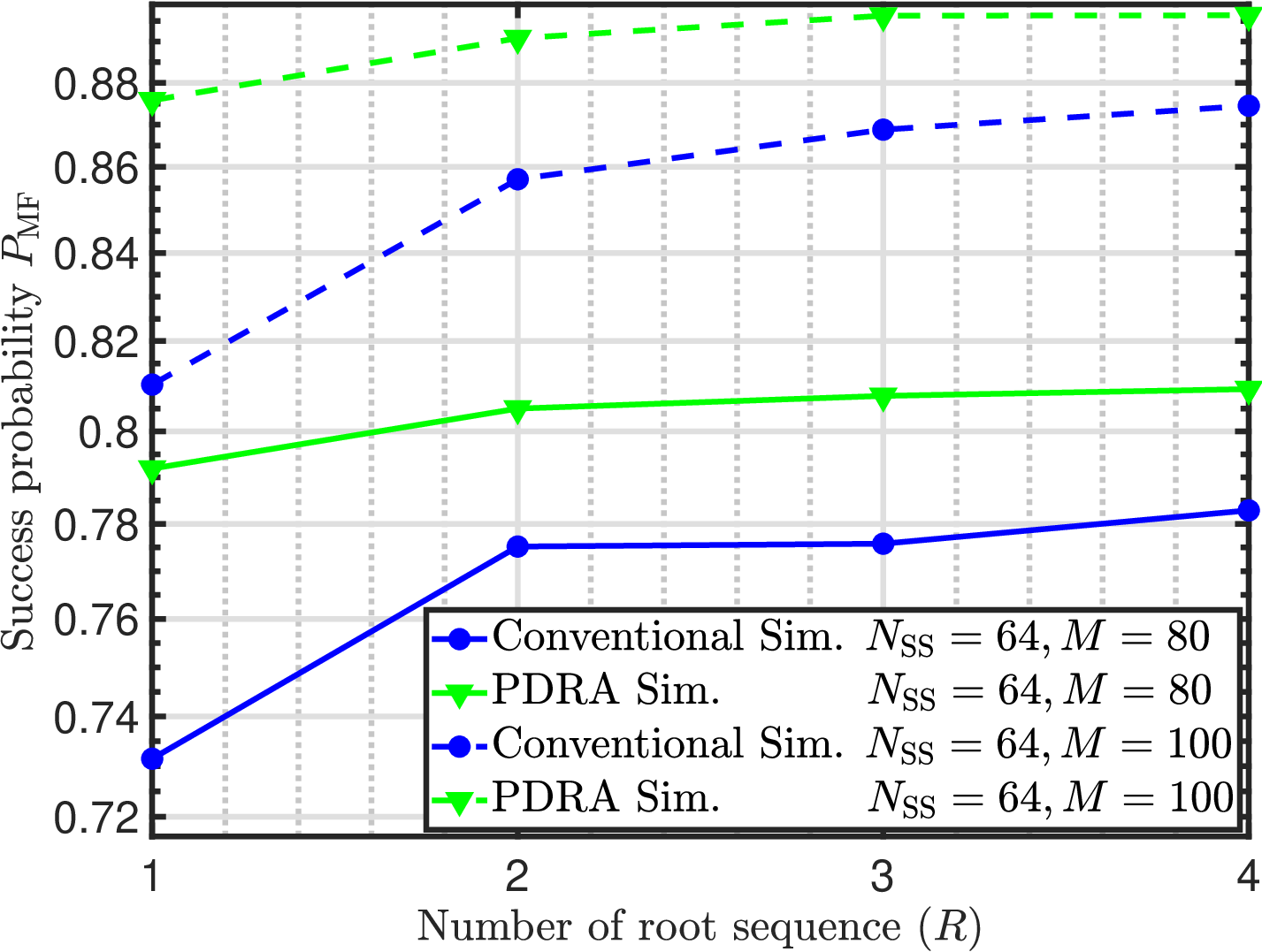} 
\label{qpsk-ber}
\label{fig-6}
\end{minipage}}
\caption{$P_{\rm{MF}}$  as a function of root number $R$  with different values of  $N_{\text{SS}}$  and $M$ under the spatially correlated Rayleigh fading channel, when    $ \rho=0.7, \alpha_{\rm{Th}}= 5\thinspace \rm{dB}$ and $ P_A=0.1\%$.
}
\end{narrow}
\end{figure}

\subsubsection{Success Probability Under A More Realistic Channel Model}

We  consider a ULA at the BS modeled with the correlation $\rho=0.7$ between adjacent antennas.
As expected,  the success probability of both schemes degrade
under the spatially correlated Rayleigh fading channel due to the channel spatial correlations among antennas as shown in Figs. \ref{fig-60} and \ref{fig-6}  [c.f. Fig. \ref{fig-3}].
  The performance degradation is lessened for BS with more  antennas since  the effect of channel spatial correlations can be mitigated in that case. 
  Figs. \ref{fig-60} and \ref{fig-6}
  also illustrate  that the proposed PDRA achieves similar performance gains over the conventional one as those  of the uncorrelated channel models 
    in  Fig. \ref{fig-3}.

\begin{rem}
For the sake of analytical tractability, we utilize the MF  as the channel estimation and data detection method 
in this work.  It is anticipated that greater performance gain can be realized via more advanced channel estimation and data detection methods, such as the linear minimum-mean-squared-error (LMMSE)  based approach.  
The purpose of this work is to shed new light on  potential directions of the pilot design for RA in massive MIMO systems, rather than on the not-so-prominent-performance-gain of the PDRA with the simple \emph{MF} receiver. The developments in this work are first geared toward insights and then toward generality.
We note that the proposed pattern-domain approach is not limited to the code-domain (exemplified by the superposition of ZC sequences) but can be extended to other orthogonal physical resources, such as time and frequency, by superimposing orthogonal building blocks in these domains (e.g., different time slots or subcarriers) to create a large set of patterns. We hope that the pattern-domain based approach may provide further impetus for innovative pilot designs for massive connectivity. 


\end{rem}

\section{Conclusion}
\label{conclu-sec}


%

This paper introduced a Pattern-Division Random Access (PDRA) scheme for massive MIMO systems, founded on a general ``superposition of the orthogonal-building-blocks"   paradigm. While demonstrated using code-domain Zadoff-Chu sequences, the core concept is broadly applicable: constructing a large set of unique access patterns by combining elementary, orthogonal resources—whether in code, time, or frequency domains.
This approach significantly expands the contention space, reducing pilot collision probability substantially without proportionally increasing physical resource overhead. Our analysis and simulations confirm that PDRA achieves notable gains over conventional schemes, even with a simple MF receiver, without excessive degradation in channel estimation or data detection.

The proposed framework paves the way for future explorations, including extensions to quasi-orthogonal building blocks, finite-field operations, and hybrid resource combinations, offering a versatile and resource-efficient pathway to support massive connectivity in beyond-5G systems.

\begin{appendices}
\section{Derivation of  ${P_{{E_0}}}$ and  ${P_{{E_1}}}$ }

For the sake of illustration clarity, we utilize     $Q$ to represent  the number of  distinct ZC  sequences in the pilot pool,
${{c}}_{i}$ to denote the ${i}$-th ZC sequence, and $M$  to represent  the number of UEs.
Consider the situation  that there are ${M}$ UEs and each UE selects  two different  ZC sequences  from the pilot pool.
It is worth noting that the selection of ${{{c}}_{1}{{c}}_{2}}$ is considered to be the  same as ${{{c}}_{2}{{c}}_{1}}$.
      The UE1 is assumed to have selected pilots ${{{c}}_{1}{{c}}_{2}}$. We divide the pilots selected by the other users into the following groups:

\begin{enumerate}

   \item The entire space is defined as a universal set ${G} = \left\{{{c_n}{c_m}\left| {1 \le n,m \le Q,n \ne m} \right.} \right\}$,  and the  number of elements of ${G}$ is expressed as $g = {{(a+1) \left({a + 2} \right)} \mathord{\left/
 {\vphantom {{(a+1) \left({a + 2} \right)} 2}} \right.
 \kern-\nulldelimiterspace} 2},$ where $a = Q - 2$.
  \item The set of pilots  with ${c_1}$ and excluding ${c_2}$ is defined as  ${A}=\left\{{{c_1}{c_n}\left| {3 \le n \le Q} \right.} \right\}$,  and the number of elements of ${A}$ is equal to $a$.
  \item The set of pilots with ${c_2}$ and excluding ${c_1}$ is defined as ${B}=\left\{{{c_2}{c_n}\left| {3 \le n \le Q} \right.} \right\}$,  and the number of elements of ${B}$ is equal to $a$.

  \item The set of pilots excluding ${c_1}$ and ${c_2}$ is defined as a complement set of the union of $A$ and $B$,  $D = {\left({A \cup B} \right) ^c} = \left\{{{c_n}{c_m}\left| {3 \le n,m \le Q,n \ne m} \right.} \right\}$,  and the number of elements of this set is expressed as ${d={a\left({a - 1} \right)} \mathord{\left/
 {\vphantom {{a\left({a - 1} \right)} 2}} \right.
 \kern-\nulldelimiterspace} 2}$.
  \item $A\cap B$ is a   singleton  containing only one element ${c_1}{c_2}$,  that is $A \cap B = \left\{{{c_1}{c_2}} \right\}$.

\end{enumerate}

\begin{figure} [t!]
\centering
\includegraphics[width=0.43\textwidth]{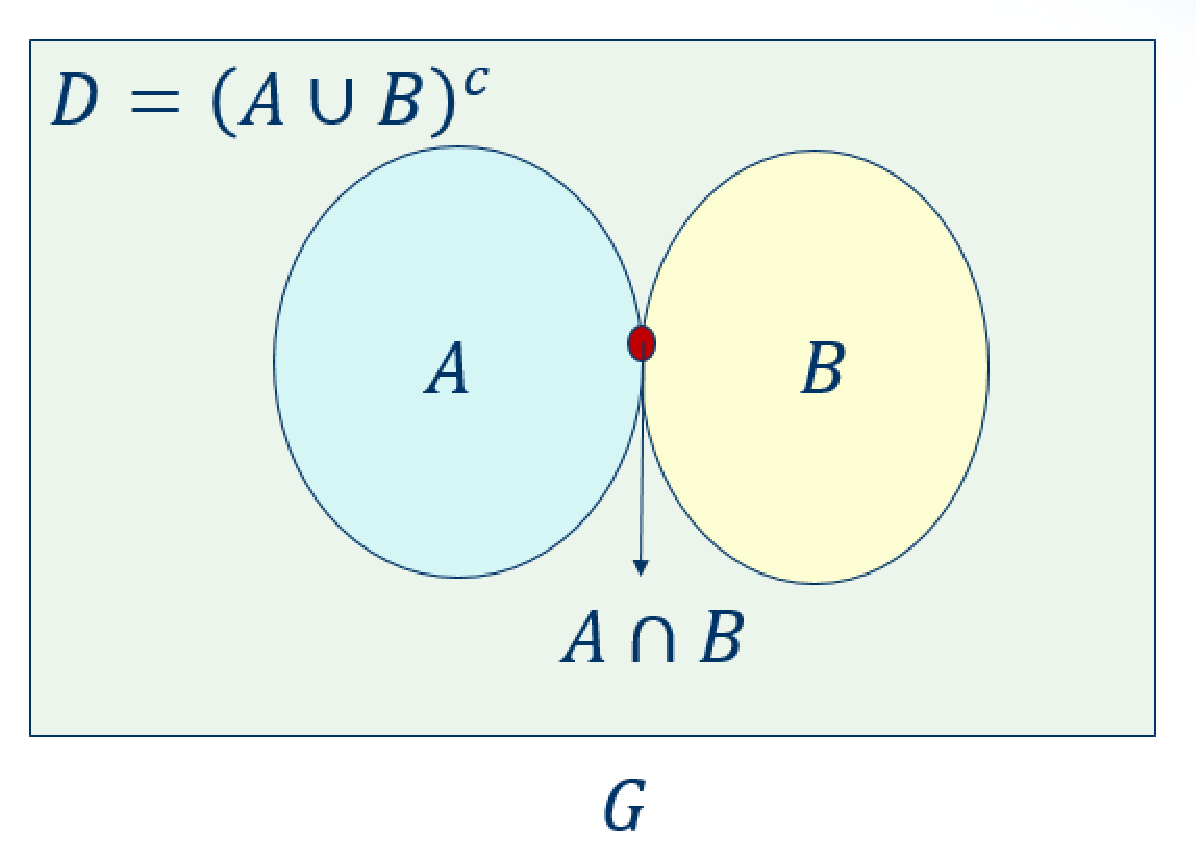}
 \caption{The groups of pilots chosen by the users.
}  \label{fig_7}
\end{figure}
The  illustration of  sets  ${A},{B},{D},$  and ${G}$ is depicted in Fig. \ref{fig_7}. ${P_{E_1}}$ is defined as probability of the event that there is only   one  pilot collision for UE1, i.e., either
 ${c_1}$ or  ${c_2}$ collides with other UEs,
which is expressed as

\begin{align}
	\label{eq20}
{P_{{E_1}}} = 2\frac{{\binom{M-1}{0}{d^0}{a^{M - 1}} +  \binom{M-1}{1}d{a^{M - 2}} +... +  \binom{M-1}{M-2}{d^{M - 2}}a}}{{{g^{\left({M - 1} \right)}}}}
 = 2{g^{- \left({M - 1} \right)}}\sum\limits_{i = 0}^{M - 2} {\binom{M-1}{i}{d^i}{a^{M - 1 - i}}}.\footnote{The discrepancy in the denominators of \eqref{p0} ($N_{PS}-1$) and \eqref{eq20} ($N_{PS}$) arises from their distinct probability spaces. \eqref{p0} is a \emph{conditional} probability within the framework of \eqref{ZF-detec-PDRA}, where the sample space for the $N-1-K$ same-root users explicitly excludes the pilot pattern already occupied by the first user. In contrast, \eqref{eq20} in Appendix A describes an \emph{unconditional} probability in a self-contained combinatorial model, where the sample space consists of \emph{all} possible pilot patterns from a single root. Thus, the denominators correctly reflect the sizes of their respective sample spaces.}
\end{align}
${P_{E_0}}$ is defined as probability of the event that there is no  pilot collision for UE1, i.e., neither
 ${c_1}$   nor    ${c_2}$ collide with other UEs, which is given by:
\begin{equation}
\label{e_0}
{P_{{E_0}}} = {\left({\frac{d}{g}} \right) ^{M - 1}}.
\end{equation}
\subsection{Note on the Denominators of Equation \eqref{p0} and Equation \eqref{eq20}}
\label{app:denominator_note}

A careful reader may note that the denominator of \eqref{p0} is $(N_{PS}-1)^{N-1-K}$, while the denominator of \eqref{eq20} is $(N_{PS})^{M-1}$. This difference is mathematically correct and intentional, stemming from the different contextual assumptions of each equation.

\begin{itemize}
    \item \textbf{Equation (12)} is derived as part of the success probability analysis in Section~III-B. It calculates the probability $P_{E_0}$ \emph{given that} $N-1-K$ other users have already selected pilot patterns from the \emph{same root sequence} as the first user. Crucially, within this conditional sample space, the specific pattern used by the first user is no longer available for selection by these same-root users. Therefore, the number of possible choices for each of these users is $N_{PS} - 1$, leading to the denominator $(N_{PS}-1)^{N-1-K}$.

    \item \textbf{Equation (20)} is presented in Appendix~A as part of a self-contained combinatorial derivation for $P_{E_1}$. This derivation considers a scenario where $M-1$ users select patterns from the entire set of patterns derived from a single root, \emph{without prior knowledge} of the first user's choice. In this unconditional probability model, the sample space must include \emph{all} possible patterns, including the one eventually chosen by the first user. Hence, the total number of choices per user is $N_{PS}$, resulting in the denominator $(N_{PS})^{M-1}$.
\end{itemize}

In summary, the denominator in \eqref{p0} represents the size of a \emph{constrained} sample space (patterns available after the first user's selection), whereas the denominator in \eqref{eq20} represents the size of the \emph{full} sample space (all possible patterns). This distinction is a standard and crucial aspect of probability theory, ensuring consistency within the logical framework of each derivation.

\section{Derivation of  $\alpha _{{\rm{MF}}}^1$}

    We first unfold  ${\bf{g}}_{1,{E_0}}^H{{\bf{h}}_1}$ and ${\bf{g}}_{1,{E_0}}^H{{\bf{h}}_n}$ as follows:
 \begin{align}
\label{e_11}
{\bf{g}}_{1,{E_0}}^H{{\bf{h}}_1}{\rm{=}}\sqrt {{P}{N_{ZC}}} {\bf{h}}_1^H{{\bf{h}}_1} + \sum\limits_{k = 2}^{K + 1} {2\sqrt {{P}} {\bf{h}}_k^H{{\bf{h}}_1}}  + {{{\bf{\tilde w}}}^H}{{\bf{h}}_1},
\end{align}

\begin{align}
\label{e_12}
{\bf{g}}_{1,{E_0}}^H{{\bf{h}}_n}{\rm{=}}\sqrt {{P}{N_{ZC}}} {\bf{h}}_1^H{{\bf{h}}_n} + \sum\limits_{k = 2}^{K + 1} {2\sqrt {{P}} {\bf{h}}_k^H{{\bf{h}}_n}}  + {{{\bf{\tilde w}}}^H}{{\bf{h}}_n}.
\end{align}
For the sake of convenience, we rewrite
(\ref{MF-est-equ})  of the manuscript  as follows: 

\begin{equation}
\label{e_5}
\begin{split}
\alpha _{{ {MF}}}^1&= \frac{{{P}{{\left| {{\bf{g}}_{1,{E_0}}^H{{\bf{h}}_1}} \right|}^{ {2}}}}}{{\sum\limits_{n = 2}^N {{P}{{\left| {{\bf{g}}_{1,{E_0}}^H{{\bf{h}}_n}} \right|}^{ {2}}} + {{\left\| {{\bf{g}}_{1,{E_0}}^H} \right\|}^2}{\sigma ^{ {2}}}}}}\\
&\mathop  \approx \limits^{\left( a \right)}  \frac{{{P}\left( {{P}{N_{ZC}}M + \sum\limits_{k = 2}^{K + 1} { {{ {4}}{P}}}  + {\sigma ^2}} \right)}}{{{P} \sum\limits_{n = 2}^N  {\left(   {{P}{N_{ZC}}  +   \sum\limits_{k = 2,k \ne n}^{K + 1}   {4{P}}   +   \sum\limits_{k = 2}^{K + 1}  {4{P}M}\delta\left(k-n\right)    +  {\sigma ^2}\;}  \right)} + \left( {{P}{N_{ZC}}  +  \sum\limits_{k = 2}^{K + 1} {   {{ {4}}{P}}}   +  {\sigma ^2}}  \right) {\sigma ^2}}}\\
&\mathop = \limits^{\left( b \right)}  \frac{{{\alpha_R ^2}{N_{ZC}}M + { {4}}{\alpha_R ^2}K{ {+}}\alpha_R}}{{{\alpha_R ^2} \left(  {N  -  1}  \right)  {N_{ZC}}  +  4{\alpha_R ^2} \left(  {K \left(  {N  -  1}  \right) -  K}  \right)   +  4{\alpha_R ^2}KM  +  \alpha_R\left(  {N  -  1}  \right) { { +}}\alpha_R {N_{ZC}}   +  { {4}}\alpha_R K  +  { {1}}}}\\
&\mathop  \approx \limits^{\left( c \right)} \frac{{{\alpha_R ^2}{N_{ZC}}M}}{{4{\alpha_R ^2}KM}} \\
&  = \frac{{{N_{ZC}}}}{{4K}}.
\end{split}
\end{equation}
 \noindent $(a) $ follows from  the fact that 1)  $\frac{{{{\left| {{\bf{h}}_n^H{{\bf{h}}_n}} \right|}^2}}}{M}\mathop  \to \limits^{M \to \infty} M$ and $\frac{{\left| {{\bf{h}}_n^H{{\bf{h}}_n}} \right|}}{M}\mathop  \to \limits^{M \to \infty} 1$;
 2)   Utilizing (\ref{e_11})  and (\ref{e_12}), we obtain
 $\frac{{{{P}{\left| {{\bf{g}}_{1,{E_0}}^H {{\bf{h}}_1}} \right|}^{\rm{2}}}}}{M}\mathop  \approx \limits^{M \to \infty}   {{P^2}{N_{ZC}}\frac{{{{\left| {{\bf{h}}_1^H {{\bf{h}}_1}} \right|}^{\rm{2}}}}}{M}{\rm{+  }}\sum\nolimits_{k = 2}^{K + 1}   {{\rm{4}}{P^2}\frac{{{{\left| {{\bf{h}}_k^H {{\bf{h}}_1}} \right|}^2}}}{M}}   +  \frac{{{P}{\sigma ^2}{{\left\| {{{\bf{h}}_1}} \right\|}^{\rm{2}}}}}{M}} $, $\frac{{{{\sigma ^2}{\left\| {{\bf{g}}_{1,{E_0}}^H} \right\|}^2}}}{M}  = 
 \frac{{{{\sigma ^2}{\left\|  {\sqrt {{P}{N_{ZC}}} {\bf{h}}_1^H  +   \sum\nolimits_{k = 2}^{K + 1}  {2\sqrt {{P}} {\bf{h}}_k^H} + {{{\bf{\tilde w}}}^H}} \right\|}^2}}}{M}$ and $\frac{{{{P}{\left| {{\bf{g}}_{1,{E_0}}^H {{\bf{h}}_n}} \right|}^2}}}{M}  \mathop  \approx \limits^{M \to \infty}   $ ${{P^2}{N_{ZC}}\frac{{{{\left| {{\bf{h}}_1^H {{\bf{h}}_n}} \right|}^2}}}{M}  +     \sum\nolimits_{k = 2}^{K + 1}   {4{P^2}\frac{{{{\left| {{\bf{h}}_k^H {{\bf{h}}_n}} \right|}^2}}}{M}}   +   \frac{{{P}{\sigma ^2}{{\left\| {{{\bf{h}}_n}} \right\|}^{\rm{2}}}}}{M}}  $;   3)  As $M \to \infty $, $\frac{{{{\left| {{\bf{h}}_k^H{{\bf{h}}_n}} \right|}^2}}}{M}$ obeys the Gamma distribution $\Gamma \left( {1,1} \right) $ when $k \ne n$.

 \noindent$(b) $
is derived based on
$\alpha_R  = P/{\sigma ^2}$.

 \noindent (c)  follows from  the fact that the non-\emph{M} part of the denominator of the third line of (\ref{e_5})   is negligible compared with those of the \emph{M}-part, as $M \to \infty $.

\end{appendices}


\begin{thebibliography}{99}
\bibitem{5billion}
 ``More than 50 billion connected devices," Stockholm,  Sweden,  Ericsson,
White Paper,  2011.
\bibitem{M2M-mag}  R.  Q.  Hu,  Y.  Qian,  H.  Chen and A.  Jamalipour,  ``Recent progress in Machine-to-Machine communications," \emph{IEEE Commun.  Mag.},  vol. 49, no. 4, pp. 24--26, Apr. 2011.
\bibitem{OptimACB}
Z. Wang and Vincent W. S. Wong,  ``Optimal access class barring for
stationary machine type communication devices with timing advance
information," \emph{IEEE Trans. Wireless Commun.},  vol. 14,  no. 10,
pp. 5374--5387,  Oct. 2015.

 \bibitem{RA-app1}
A. Laya,  L. Alonso,  and J. Alonso-Zarate,  ``Is the random access channel of LTE and LTE-A suitable for M2M communications? A survey
of alternatives," \emph{IEEE Commun. Surveys Tuts}.,  vol. 16,  no. 1,  pp. 4--16,
first Quart. 2014.
 \bibitem{WeiRAC}
C. Wei, R. Cheng and S. Tsao, ``Modeling and estimation of one-shot random access for finite-user multichannel slotted ALOHA systems,"  \emph{IEEE Communi. Lett.}, vol. 16, no. 8, pp. 1196--1199, Aug. 2012.
 \bibitem{GroupRAC}
K. Lee \emph{et al}.,  ``A group-based communication scheme based on the
location information of MTC devices in cellular networks," in \emph{Proc}. \emph{IEEE Int. Conf. Commun}.  (\emph{ICC}),  Ottawa,  ON,  Canada,  Jun. 2012, pp. 4899--4903.
\bibitem{SlotMAC}
 3GPP R2--103759 (2010).  Load distribution for MTC devices,  ALU/ASB,  RAN270-bis.

\bibitem{ACB-3gpp}
  ``Study on RAN improvements for machine-type communications,"
3GPP,  Sophia Antipolis,  France,  Tech. Rep. 37.868 V14.0.0,  Sep.  2017.
  \bibitem{PRADA} T.-M. Lin,  C.-H. Lee,  J.-P. Cheng,  and W.-T. Chen,  ``PRADA: Prioritized
random access with dynamic access barring for MTC in 3GPP LTE-A
networks," \emph{IEEE Trans. Veh. Technol}.,  vol. 63,  no. 5,  pp. 2467--472,
Jun.  2014.
  \bibitem{CRMA-ambiguity}
S. Vural, N. Wang, G. Foster and R. Tafazolli,  ``Success probability of multiple-preamble-based single-attempt random access to mobile networks," in \emph{IEEE Communi. Letters}, vol. 21, no. 8, pp. 1755--1758, Aug. 2017.  
\bibitem{Success-RA-MIMO} 
J. Ding, D. Qu, H. Jiang and T. Jiang, ``Success probability of grant-free random access with massive MIMO,"  \emph{IEEE Internet of Things Journal}, vol. 6, no. 1, pp. 506--516, Feb. 2019.  
\bibitem{NOMA-RA} L. Liu and W. Yu, ``Massive connectivity with massive MIMO Part I:
Device activity detection and channel estimation," \emph{IEEE Trans. Signal
Process}., vol. 66, no. 11, pp. 2933--2946, Jun. 2018.
\bibitem{TwoStep-RA}
K. Senel and E. G. Larsson, ``Grant-free massive MTC-enabled massive MIMO: a compressive sensing approach,"  \emph{IEEE Trans.   Communi.}, vol. 66, no. 12, pp. 6164--6175, Dec. 2018.  
\bibitem{RAC-3GPP}
Evolved Universal Terrestrial Radio Access (E-UTRA) ; Medium Access
Control (MAC)  Protocol Specification; Release 13,  3GPP TS 36. 321,
2017.

\bibitem{PRACH}  S. Sesia, M. Baker, and I. Toufik, \emph{LTE-the UMTS Long Term Evolution:
from Theory to Practice}.  John Wiley $\&$ Sons, 2011.
\bibitem{PC-3GPP}
Technical Specification Group Radio Access Network, NR, Physical Layer Procedures, TS 36.213 V15.0.0, Dec. 2017.

 \bibitem{Spatial Channel Model-3gpp}
 ``Spatial Channel Model for Multiple Input Multiple Output (MIMO)
Simulations (Release 13) ", document TR 25. 996, 3GPP, Dec.  2019.
\end{thebibliography}
\end{document}